# Security Analysis of Vehicular Ad Hoc Networks (VANET)

Ghassan Samara[#1], Wafaa A.H. Al-Salihy[*2], R. Sures[#3]

[#]National Advanced IPv6 Center, Universiti Sains Malaysia
Penang, Malaysia
[1]ghassan@nav6.org, [3]sures@nav6.org

[*]School of Computer Science, Universiti Sains Malaysia
Penang, Malaysia
[2]wafaa@cs.usm.my

**Abstract-** Vehicular Ad Hoc Networks (VANET) has mostly gained the attention of today's research efforts, while current solutions to achieve secure VANET, to protect the network from adversary and attacks still not enough, trying to reach a satisfactory level, for the driver and manufacturer to achieve safety of life and infotainment. The need for a robust VANET networks is strongly dependent on their security and privacy features, which will be discussed in this paper.
In this paper a various types of security problems and challenges of VANET been analyzed and discussed; we also discuss a set of solutions presented to solve these challenges and problems.

## I. INTRODUCTION

Recent year's rapid development in wireless communication networks has made Inter-Vehicular Communications (IVC) and Road-Vehicle Communications (RVC) possible in Mobile Ad Hoc Networks (MANETs), this has given birth to a new type of MANET known as the Vehicular Ad Hoc Network (VANET), aiming to enable road safety, efficient driving, and infotainment.

The world today is living a combat, and the battle field lies on the roads, the estimated number of deaths is about 1.2 million people yearly worldwide [15], and injures about forty times of this number, without forgetting that traffic congestion that makes a huge waste of time and fuel [1].

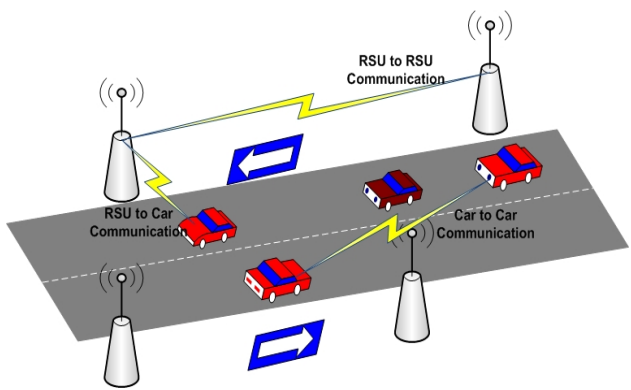

Fig. 1 VANET Structure

Vehicular Ad hoc Networks (VANET) is part of Mobile Ad Hoc Networks (MANET), this means that every node can move freely within the network coverage and stay connected, each node can communicate with other nodes in single hop or multi hop, and any node could be Vehicle, Road Side Unit (RSU).

In the year 1998, the team of engineers from Delphi Delco Electronics System and IBM Corporation proposed a network vehicle concept aimed at providing a wide range of applications [14]. With the advancements in wireless communications technology, the concept of network car has attracted the attention all over the world.

In recent years, many new projects have been launched, targeting on realizing the dream of networking car and successful implementation of vehicular networks. The project Network On Wheels (NOW) [3] is a German research project founded by DaimlerChrysler AG, BMW AG, Volkswagen AG, Fraunhofer Institute for Open Communication Systems, NEC Deutschland GmbH and Siemens AG in 2004, The project adopts an IEEE 802.11 standard for wireless access, The main objectives of this project are to solve technical issues related to communication protocols and data security for car-to-car communications. The Car2Car Communication Consortium [16] is initiated by six European car manufacturers. Its goal is to create a European industrial standard for car-to-car communications extend across all brands. FleetNet [16] was another European program which ran from 2000 to 2003 this ad hoc research was dominated by efforts to standardize MANET protocols, and this MANET research focused on the network layer[2], the ultimate challenge was to solve the problem of how to reach nodes not directly within radio range by employing neighbors as forwarders, while the European Commission is pushing for a new research effort in this area in order to reach the goal of reducing the car accidents of 50% by 2010, aiming to reach a satisfactory level of secure VANET.

The radio used for the communication is Dedicated Short Range Communications (DSRC), which been allocated as new band in 1999 by the Federal Communications Commission (FCC)[3], the band allocated was 75 MHz at 5.9 GHz frequency for Intelligent Transport System (ITS) applications in north America.

VANET security should satisfy four goals, it should ensure that the information received is correct (information authenticity), the source is who he claims to be (message integrity and source authentication), the node sending the message cannot be identified and tracked (privacy) and the system is robust.

Our paper presents in section 2 an analysis of VANET attack and attackers to show the problems that VANET facing, in section 3 we analyzed VANET challenges like mobility and privacy which considered the hardest security problems of VANET, in section 4 we list the security requirement that must exist to achieve the security system, in section 5 we discussed the current solution for the challenges and attacks and requirement to achieve a secure system, that been addressed by other papers and researchers.

### a. HOW VANET WORKS

Vehicular Networks System consists of large number of nodes, approximately number of vehicles exceeding 750 million in the world today [4], these vehicles will require an authority to govern it, each vehicle can communicate with other vehicles using short radio signals DSRC (5.9 GHz), for range can reach 1 KM, this communication is an Ad Hoc communication that means each connected node can move freely, no wires required, the routers used called Road Side Unit (RSU), the RSU works as a router between the vehicles on the road and connected to other network devices.

Each vehicle has OBU (on board unit), this unit connects the vehicle with RSU via DSRC radios, and another device is TPD (Tamper Proof Device), this device holding the vehicle secrets, all the information about the vehicle like keys, drivers identity, trip details, speed, rout …etc, see figure 1.

### II. VANET SECURITY CONCERNS

VANET suffer from various attacks; these attacks are discussed in the following subsections:

#### A. *ATTACKS*

In this paper we are concentrating on attacks perpetrated against the message itself rather than the vehicle, as physical security is not in the scope of this paper.

*1) Denial of Service attack,*
This attack happens when the attacker takes control of a vehicle's resources or jams the communication channel used by the Vehicular Network, so it prevents critical information from arriving. It also increases the danger to the driver, if it has to depend on the application's information.

For instance, if a malicious wants to create a massive pile up on the highway, it can make an accident and use the DoS attack to prevent the warning from reaching to the approaching vehicles [1], [5], [6], and [7]. See figure 2.

Authors in [1] discussed a solution for DoS problem and saying that the existing solutions such as hopping do not completely solve the problem, the use of multiple radio transceivers, operating in disjoint frequency bands, can be a feasible approach but even this solution will require adding new and more equipments to the vehicles, and this will need more funds and more space in the vehicle. The authors in [12], proposed a solution by switching between different channels or even communication technologies (e.g., DSRC, UTRA-TDD, or even Bluetooth for very short ranges), if they are available, when one of them (typically DSRC) is brought down.

*2) Message Suppression Attack,*
An attacker selectively dropping packets from the network, these packets may hold critical information for the receiver, the attacker suppress these packets and can use them again in other time[5].

The goal of such an attacker would be to prevent registration and insurance authorities from learning about collisions involving his vehicle and/or to avoid delivering collision reports to roadside access points [17].

For instance, an attacker may suppress a congestion warning, and use it in another time, so vehicles will not receive the warning and forced to wait in the traffic.

*3) Fabrication Attack,*
An attacker can make this attack by transmitting false information into the network, the information could be false or the transmitter could claim that it is somebody else.

This attack includes fabricate messages, warnings, certificates, identities [5], [7] [17].

*4) Alteration Attack,*
This attack happens when attacker alters an existing data, it includes delaying the transmission of the information, replaying earlier transmission, or altering the actual entry of the data transmitted [5].

For instance, an attacker can alter a message telling other vehicles that the current road is clear while the road is congested [17].

*5) Replay Attack,*
This attack happens when an attacker replay the transmission of an earlier information to take advantage of the situation of the message at time of sending [5].

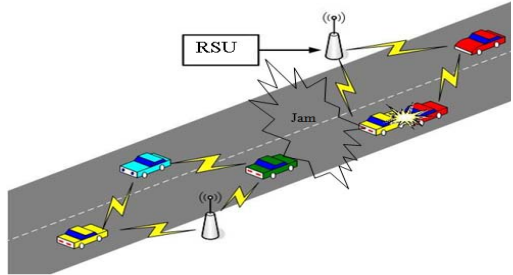

Fig. 2 DoS Attack

Basic 802.11 security has no protection against replay. It does not contain sequence numbers or timestamps. Because of keys can be reused, it is possible to replay stored messages with the same key without detection to insert bogus messages into the system. Individual packets must be authenticated, not just encrypted. Packets must have timestamps.

The goal of such an attack would be to confuse the authorities and possibly prevent identification of vehicles in hit-and-run incidents [17].

*6) Sybil Attack,*

This attack happens when an attacker creates large number of pseudonymous, and claims or acts like it is more than a hundred vehicles, to tell other vehicles that there is jam ahead, and force them to take alternate route[5],[8]. See Figure 3.

Sybil attack *depends on how cheaply identities can be generated, the degree to which the system accepts inputs from entities that do not have a chain of trust linking them to a trusted entity, and whether the system treats all entities identically.*

For instance an attacker can pretend and act like a hundred vehicle to convince the other vehicles in the road that there is congestion, go to another rout, so the road will be clear.

### B. ATTACKERS

*1) Selfish Driver,*

The general idea for trust in Vehicular Network is that all vehicles must be trusted initially, these vehicles are trusted to follow the protocols specified by the application, some drivers try to maximize their profit from the network, regardless the cost for the system by taking advantage of the network resources illegally [5].

A Selfish Driver can tell other vehicles that there is congestion in the road, so they must choose an alternate route, so the road will be clear for it. See figure 4.

*2) Malicious Attacker,*

This kind of attacker tries to cause damage via the applications available on the vehicular network. In many cases, these attackers will have specific targets, and they will have access to the resources of the network [1], [5].

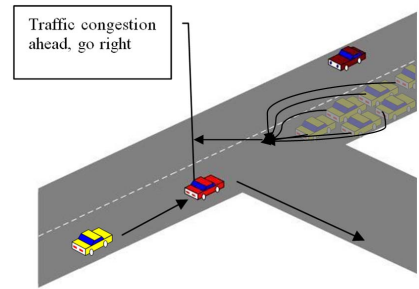

Fig. 3 Sybil Attack.

For instance, a terrorist can issue a deceleration warning, to make the road congested before detonating a bomb.

*3) Pranksters,*

Include bored people probing for vulnerabilities and hackers seeking to reach fame via their damage [5].

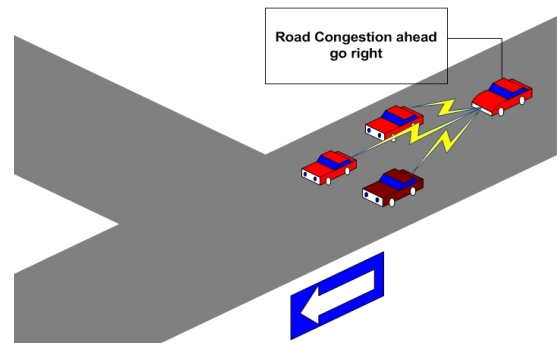

Fig. 4 Selfish Driver

For instance, a prankster can convince one vehicle to slow down, and tell the vehicle behind it to increase the speed.

### III. VEHECULAR NETWORKS CHALENGES

*1) Mobility*

The basic idea from Ad Hoc Networks is that each node in the network is mobile, and can move from one place to another within the coverage area, but still the mobility is limited, in Vehicular Ad Hoc Networks nodes moving in high mobility, vehicles make connection throw their way with another vehicles that maybe never faced before, and this connection lasts for only few seconds as each vehicle goes in its direction, and these two vehicles may never meet again. So securing mobility challenge is hard problem.

There is many researches have addressed this challenge [5], [9], but still this problem unresolved.

*2) Volatility*

The connectivity among nodes can be highly ephemeral, and maybe will not happen again, vehicles travelling throw coverage area and making connection with other vehicles, these connections will be lost as each car has a high mobility, and maybe will travel in opposite direction[1],[5].

Vehicular networks lacks the relatively long life context, so personal contact of user's device to a hot spot will require long life password and this will be impractical for securing VC.

*3) Privacy VS Authentication*

The importance of authentication in Vehicular Ad Hoc Networks is to prevent Sybil Attack that been discussed earlier [8].

To avoid this problem we can give a specific identity for every vehicle, but this solution will not be appropriate for the most of the drivers who wish to keep their information protected and private[1],[5].

*4) Privacy VS Liability*

Liability will give a good opportunity for legal investigation and this data can't be denied (in case of accidents)[1], in other hand the privacy mustn't be violated and each driver must have the ability to keep his personal information from others (Identity, Driving Path, Account Number for toll Collector etc.).

*5) Network Scalability*

The scale of this network in the world approximately exceeding the 750 million nodes [4], and this number is growing, another problem arise when we must know that there is no a global authority govern the standards for this network [1], [5], [7], for example: the standards for DSRC in North America is deferent from the DSRC standards in Europe, the standards for the GM Vehicles is deferent from the BMW one.

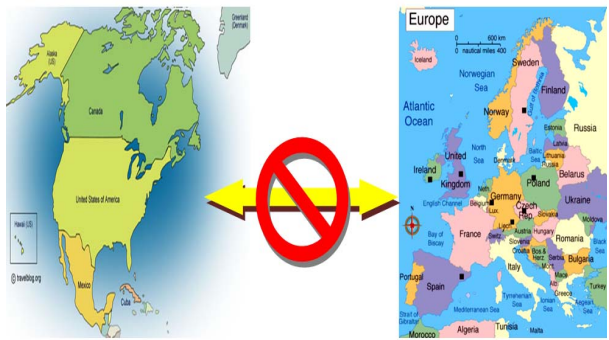

Fig. 5: Europe VS North America Standards.

*6) Bootstrap*

At this moment only few number of cars will be have the equipment required for the DSRC radios, so if we make a communication we have to assume that there is a limited number of cars that will receive the communication, in the future we must concentrate on getting the number higher, to get a financial benefit that will courage the commercial firms to invest in this technology [5].

## IV. SECURITY REQUIREMENTS

*1- Authentication*

In Vehicular Communication every message must be authenticated, to make sure for its origin and to control authorization level of the vehicles, to do this vehicles will assign every message with their private key along with its certificate, at the receiver side, the receiver will receive the message and check for the key and certificate once this is done, the receiver verifies the message [1], [5].

Signing each message with this, causes an overhead, to reduce this overhead we can use the approach ECC (Elliptic Curve Cryptography), the efficient public key cryptosystem, or we can sign the key just for the critical messages only.

*2. Availability*

Vehicular network must be available all the time, for many applications vehicular networks will require real-time, these applications need faster response from sensor networks or even Ad Hoc Network, a delay in seconds for some applications will make the message meaningless and maybe the result will be devastating[1][5].

Attempting to meet real-time demands makes the system vulnerable to the DoS attack. In some messages, a delay in millisecond makes the message meaningless; the problem is much bigger, where the application layer is unreliable, since the potential way to recover with unreliable transmission is to store partial messages in hopes to be completed in next transmission.

*3. Non-repudiation*

Non-repudiation will facilitate the ability to identify the attackers even after the attack happens [5], [8]. This prevents cheaters from denying their crimes.

Any information related to the car like: the trip rout, speed, time, any violation will be stored in the TPD, any official side holding authorization can retrieve this data.

*4. Privacy*

Keeping the information of the drivers away from unauthorized observers, this information like real identity, trip path, speed etc…

The privacy could be achieved by using temporary (anonymous) keys, these keys will be changed frequently as each key could be used just for one time and expires after usage [1], all the keys will be stored in the TPD, and will be reloaded again in next time that the vehicle makes an official checkup [5].

For preserving the real identity of the driver, an ELP (Electronic License Plate) is used, this license is installed in the factory for every new vehicle, it will provide an identification number for the vehicle, to identify the vehicle in anywhere, with the RFID technology to hold the ELP.

In case when the police or any official wants the real identity, it can take an order from the judge to recover the identity of specific vehicles ELP.

### 5. Real-time constraints

Vehicles move in high speed, this will require a real-time response in some situation, or the result will be devastating [5].

Current plans for vehicular networks rely on the emerging standard for dedicated short-range communications (DSRC), based on an extension to the IEEE 802.11 technology.

### 6. Integrity

Integrity for all messages should be protected to prevent attackers from altering them, and message contents to be trusted [1].

### 7. Confidentiality

The privacy of each driver must be protected; the messages should be encrypted to prevent outsiders from gaining the drivers information [1].

## V. CURRENT PROPOSED SOLUTIONS

In VANET many security solutions been proposed, and large number of papers were introduced to solve the above problems, the authors in [1] and in [7] suggested the use of VPKI (Vehicular Public Key Infrastructure) as a solution, where each node will have a public/private key. When a vehicle sends a safety message, it signs it with its own private key and adds the Certificate Authority (CA's) certificate as follows:

V → r: M, SigPrKV [M|T], Certv [7]

Where V is the sending vehicle, r represents the message receivers, M is the message, | is the concatenation operator, and T is the timestamp to ensure message freshness (it can be obtained from the security device). The receivers of the message will obtain the public key of V using the certificate and then verify V's signature using its certified public key. In order to do this, the receiver should have the public key of the CA [12]; this solution is cited in [3], [5], [10], and [11].

Authors in [18] suggested an idea of using the group signature, but this idea has a major drawback that it is causing a great overhead, every time that any vehicle enters the group area, the group public key and the vehicle session key for each vehicle that belongs to the group must be changed and transmitted, another issue must be considered that the mobility of the VANET prevents the network from making a static group, so the group is changing all the time, and the signatures and keys frequently changed and transmitted, group signature is also mentioned in [10][19], as the authors proposed a protocol for guarantee the requirements of the security and privacy, and to provide the desired traceability and liability, but the result of the study was not quit encouraging, After 9 ms for group signature verification delay, the average message loss ratio was 45%, another result was the loss ratio reaches as high as 68% when the traffic load is 150 vehicles .

The other solution been suggested is the use of CA and this requires infrastructure for it. VANET requires a large number of CA to govern it. until now we don't have a real authority that govern the world of VANET, the CA been suggested by[4],[7],[10],[11],[12],[13] , all of these researchers mentioned the CA to handle all the operations of certificate : generating, renewing and revoking, and CA must be responsible in initiating keys, storing, managing and broadcasting the CRL.

Authors in [1] also discussed how to maintain the authentication for the message, where vehicles will sign each message with their private key and attach the corresponding certificate. Thus, when another vehicle receives this message, it verifies the key used to sign the message and if everything is correct, it verifies the message, and they have proposed the use of ECC to reduce the overhead as mentioned before in section 4.1, while authors in [3] suggested another way to use the keys, by using short term certificates and long term, long term certificates are used for authentication while short term certificates are used for data transmission using public/private key cryptography. Safety messages are not encrypted as they are intended for broadcasting, but their validity must be checked; therefore a source signs a message and sends it without encryption with its certificate; other nodes receiving the message validate it using the certificate and signature and may forward it without modification if it is a valid message, so any adversary can inject false information as a safety message, as it doesn't to be encrypted, it also can steal the certificate from any other safety message and send unencrypted message contains false information along with the stolen certificate claiming that the safety message originated from another vehicle.

Using VPKI in VANET accompanied with some challenges, like certificate of an attacker that must be revoked, authors in[1] discussed the Certificate Revocation solution, this solution is used to revoke the expired certificate to make other vehicles aware of their invalidity, and The most common way to revoke certificates is the distribution of CRLs (Certificate Revocation Lists) that contains all revoked certificates, but this method has some drawbacks: First, CRLs can be very long due to the enormous number of vehicles and their high mobility. Second, the short lifetime of certificates still creates a vulnerability window and last one is that there is no infrastructure for the CRL. It is also mentioned some protocols for revocations like RTPD (Revocation Protocol of the Tamper-Proof Device), RCCRL (Revocation protocol using Compressed Certificate Revocation Lists), and DRP (Distributed Revocation Protocol), these protocols also been discussed in details in [4], and been proposed in [11], saying that the use of CRL is not appropriate anymore and these protocols are better, but these methods rely on monitoring, so every vehicle has to monitor and detect all the vehicles around it, but this method didn't consider the reputation system, as it is a possible for number of adversary vehicles to make an accusation and causing of an unnecessary revocation, the best result obtained from DRP simulation is that just 25% if the current road vehicles will receive the warning, which is too low.

Authors in [1] mentioned a solution that will help to maintain the privacy by using a set of anonymous keys that change frequently (every couple of minutes) according to the driving speed. Each key can be used only once and expires after its usage; only one key can be used at a time. These keys are preloaded in the vehicle's TPD for a long duration; each key is certified by the issuing CA and has a short lifetime (e.g., a specific week of the year). In addition, it can be traced back to the real identity of the vehicle ELP, the drawback of this solution that the keys need storage.

In[3] authors mentioned that In the IEEE WAVE standard vehicles can change their IP addresses and use random MAC addresses to achieve security, IP version 6 has been proposed for use in vehicular networks. Cars should be able to change their IP addresses so that they are not traceable, however it is not clear how this will be achieved.

Moreover this can cause inefficiency in address usage since when a new address is assigned the old address cannot be reused immediately. Delayed packets will be dropped when the car changes its IP address which causes unnecessary retransmissions.

Authors in [5] added another proposed solution by making regular inspections, where in most U.S. states all vehicles must pass inspection once a year. This yearly trip to the mechanic provides interesting possibilities for security maintenance in addition to the typical maintenance to be performed.

## VI. CONCLUSIONS AND FUTURE WORK

Vehicular Ad Hoc Networks is promising technology, which gives abundant chances for attackers, who will try to challenge the network with their malicious attacks. This paper gave a wide analysis for the current challenges and solutions, and critics for these solutions, in our future work we will propose new solutions that will help to maintain a securer VANET network, and test it by simulation.